\documentstyle{article}
\catcode`\@=11

\@addtoreset{equation}{section}
\catcode`\@=12
\input{amssym}
\def\sq{\square}
\def\bsq{\blacksquare}
\def\mod{\mathop{\rm mod\,}\nolimits}
\font\bbb=msbm10
\def\C{\hbox{\bbb C}}
\def\F{\hbox{\bbb F}}
\def\N{\hbox{\bbb N}}
\def\Z{\hbox{\bbb Z}}

\begin{document}

\author{A. K. Pogrebkov \\ Steklov Mathematical Institute, \\
Gubkin str. 8, Moscow 117966, GSP-1, \\
RUSSIA \\ e-mail: pogreb@mi.ras.ru}
\title{Discrete Schr\"{o}dinger equation on finite field and associated 
cellular automaton}
\maketitle

\begin{abstract}
The discrete Schr\"{o}dinger equation with potential belonging to 
$\F_{2}$ is solved explicitly. On this base the associated 
($1+1$)-dimensional cellular automaton is examined and corresponding set of
integrals of motions is constructed.
\end{abstract}

\section{Introduction}

Cellular automata are dynamical systems in the discrete space--time with
values belonging to some finite field. For example, in the 
simplest situation the dependent variable $q_{n}^{t}$ takes values $0$ and 
$1$, i.e., belongs to $\F_{2}$ for any $n,t\in {\Z}$, where $n$ and $t$
are considered as space and time variables correspondingly.
Initial data  are given 
as a set of zeroes and ones numbered by $n\in\Z$ and time evolution is a law 
that enables us to construct $q_{n}^{t+1}$ also as a set of zeroes and ones 
if $q_{n}^{t}$ is known. Cellular automata attract essential interest in the 
literature because of the wide range of their applications in different 
sciences, from physics to biology, from chemistry to social sciences. 
Detailed references for these and many other applications can be found 
in~[1--5]. These automata are also subject to intensive mathematical 
study, see for example~[6--18], but nevertheless the 
mathematical theory of cellular automata is far from completeness. In some 
cases (see~[4]) these dynamical system demonstrate stochastic behavior, 
in other cases they behave as solitonic systems~[7--18]. The situation 
becomes even more intriqueing if to take into account that, at least, two 
families of the cellular automata have Lax pairs. Nevertheless, the problem 
of integrability of such cellular automata is still opened as these Lax 
representation are satisfied only on corresponding finite fields as well. 

Above mentioned families of the cellular automata that have associated Lax 
representations were considered in~[12--15]. All of them are of so 
called filter type~[6]. In works~[12,15] these were automata with 
Lax operator $L$ to be the known Ablowitz--Ladik system~[19]. In 
work~[13] $L$ was chosen to be the discrete (1-dimensional) Schr\"{o}dinger 
operator, that is also known in the literature~[20,21]. As was mentioned 
in~[12,13,15] (and also discussed below) known results on either 
Ablowitz--Ladik, or discrete Schr\"{o}dinger equation cannot be applied to 
the case where potential belongs to a finite field as in this case some 
of the Jost solutions do not exist. Correspondingly, this unable us to
introduce the spectral data, so the standard scheme~[22--24] of the 
inverse spectral transform fails for this kind of problems requires essential
modification. 

In this note we demonstrate that, on the other side, in consideration of the 
discrete problems some essential simplifications occur. In particular we 
show that the Jost solutions of the discrete Schr\"{o}dinger equation (the 
existing ones!) can be constructed explicitly. We apply result of 
this construction for the simplest example of the cellular automata 
introduced in~[13], 
\begin{equation}
q_{n}^{t+1}=q_{n}^{t}+q_{n-2}^{t+1}q_{n+1}^{t}+q_{n-1}^{t+1}
q_{m+2}^{t},\qquad \mod 2.  \label{a50}
\end{equation}
Here dependent variable $q_{n}^{t}\in\{0,1\}$ for any values of two 
independent variables $n,t\in {\Z}$. Lax representation for this equation 
is given in [13] as 
\begin{equation}
L_{}^{t+1}A=AL_{}^{t},\qquad \mod 2,  \label{a57}
\end{equation}
where matrices $L_{m,n}^{t}$ and $A_{m,n}^{}$ are equal to
\begin{eqnarray}
&&L_{m,n}^{t}(w)=\delta _{m,n+1}^{}+(1+q_{m}^{t})\delta _{m,n-1}^{}- 
\biggl(w+\frac{1}{w} \biggr)\delta _{m,n},  \label{a5}\\
&&A_{m,n}^{}=\delta _{m,n}^{}-\delta _{m+2,n}^{}q_{m-1}^{t+1}q_{m+2}^{t}.
\label{a58}
\end{eqnarray}
It was also shown in~[13] that if we consider $q_n^t$ at some moment of 
$t$ to be finitely supported then $q_n^{t+1}$ and so on also have finite 
support. More exactly, let us introduce support of sequence $q_n^{t}$  at 
moment $t$,
\begin{equation}
{\cal K}_{}^{t}\equiv \{k_{1}^{t}<\ldots <k_{N^{t}}^{t}\}=\{k\in \hbox{\Z}
|q_{k}^{t}=1\},  \label{a2}
\end{equation}
so that $k_{1}^{t}$ and $k_{N^{t}}^{t}$ are bottom and upper borders of the 
support and $N^{t}<\infty$. Then borders of the support are preserved under 
time evolution,
\begin{equation}
k_{1}^{t+1}=k_{1}^{t},\qquad k_{N^{t+1}}^{t+1}=k_{N^{t}}^{t},  \label{a63}
\end{equation}
and, moreover, support ${\cal K}^{t}$ is decomposed into 
``islands"---subsets of support separated by three or more consequent 
sites occupied by zeros. These islands evolve independently, 
in particular, their borders are 
also integrals of motion. Simple example of time evolution of 
cellular automata~(\ref{a50}) is presented in Fig.\ 1, where empty 
boxes correspond 
to zero sites and black boxes correspond to the sites occupied by units. We 
presented here the case of two islands. It is clear that time evolution is 
reduced to the motion of single and double defects inside islands, and that 
it is enough to consider the case where ${\cal K}^{t}$ is a single island. 
Finally, let us notice that equation~(\ref {a50}) is invariant with respect 
to the substitution $q_n^t\to q_{-n}^{-t+1}$, so this cellular automaton is 
time reversible. 

Spectral problem for the operator~(\ref{a5}),
\begin{equation}
\sum_{n}L_{m,n}^{t}(w)\psi _{n}^{t}(w)=0,  \label{6}
\end{equation}
here also has to be understood modulo 2. This means that for any $n$ and $t$ 
solution $\psi _{n}^{t}(w)$ is formal Laurent series in $w\in\C$ with 
coefficients belonging to $\F_2$. Omitting for simplicity subscript $t$ we 
rewrite~(\ref {6}) by means of~(\ref {a5}) as
\begin{equation}
\psi _{m-1}^{}+(1+q_{m}^{})\psi _{m+1}^{}=\biggl(w+\frac{1}{w}\biggr)
\psi _{m}^{},\qquad \mod 2.  \label{1}
\end{equation}
This is known discretization of the famous Schr\"{o}dinger equation
and for generic $q_m$ it can be studied in analogy with the standard 
Schr\"{o}dinger equation, see~[20,21]. We already mentioned 
that it is just modulo 2 condition 
that prevents from the use of the corresponding results of~[20,21] 
here. Indeed, for all $m\notin{\cal K}$ (i.e., $q_m=0$) we can determine 
$\psi_{m+1}$ if $\psi_{m}$ and $\psi_{m-1}$ are known. On the other side, if 
$m\in{\cal K}$ (i.e., $q_m=1$), then $\psi_{m+1}$ drops out from~(\ref {1}) 
and $\psi_{m}$ and $\psi_{m-1}$ cannot be chosen independently. This means 
that in all cases where ${\cal K}$ is not empty equation~(\ref {1}) can be 
solved by means of the sweeping from the left, but cannot be swept from the 
right.

\section{Solution of the discrete Schr\"{o}dinger\protect\newline equation}

In order to construct solutions of the discrete Schr\"{o}dinger equation let 
us mention first that in the case where $q_{m}\in \F_{2}$ we have
\begin{equation}
(1+q_{m}^{})(\mod 2)=1-q_{m}^{},  \label{3}
\end{equation}
so instead of (\ref{1}) we start with construction of solutions of the
infinite system 
\begin{equation}
\psi _{m-1}^{}+(1-q_{m}^{})\psi _{m+1}^{}=\biggl(w+\frac{1}{w}\biggr) \psi
_{m}^{},  \label{4}
\end{equation}
where $q_{m}\in\{0,1\}$ for any $m\in\Z$, $w\in \C$ and condition 
that~(\ref{4}) is satisfied by $\mod 2$ is omitted. It is clear that such 
solutions of~(\ref{4}) resolves~(\ref {1}) in the sense mentioned in 
Introduction. 

If $m\leq k_1$ or $m\geq k_N+1$ (see~(\ref{a2})) we have two 
explicit solutions of~(\ref{4}): $w^{m}$ and $w^{-m}$. Thus for such $m$ 
general solution of (\ref{4}) is given as linear combination of these two 
solutions. Let
\begin{equation}
x_{m}^{}(w)=w_{}^{m}\psi _{m}^{}(w),  \label{7}
\end{equation}
that due to (\ref{4}) obeys equation 
\begin{equation}
x_{m-1}^{}-(1+z)x_{m}^{}+z(1-q_{m}^{})x_{m+1}^{}=0,  \label{8}
\end{equation}
where for simplicity we denoted
\begin{equation}
z=w_{}^{-2}.  \label{503}
\end{equation}
In what follows we use the same notations, say $x_m(z)$ or $x_m(w)$, for 
functions of either $z$, or $w$, related by means of this substitution. 
It is clear that if $\psi _{m}(w)$ solves (\ref{4}) then 
\begin{equation}
\tilde{\psi}_{m}^{}(w)=\psi _{m}^{}(w_{}^{-1})  \label{a11}
\end{equation}
gives another solution of (\ref{4}). Let us introduce $\tilde{x}_{m}^{}$ in 
analogy with (\ref{7}), then 
\begin{equation}
\tilde{x}_{m}^{}(z)=z_{}^{-m}x_{m}^{}(z_{}^{-1})  \label{13}
\end{equation}
gives another solution of~(\ref{8}).

Mentioned in Introduction problem of the study of the spectrum~(\ref{8}) is 
reflected in the existence of the trivial solution 
\begin{equation}
x_{0,m}^{}(z)=\left\{\begin{array}{ll}
\displaystyle{\frac{1-z^{k_{N}-m}}{1-1/z}},&\qquad m\geq k_{N}^{}, \\ 
0,&\qquad m\leq k_{N}^{}.
\end{array} \right.  \label{10}
\end{equation}
This solution is invariant (up to $m$-independent factor) with respect to 
transformation~(\ref {13}).

The Jost solution is defined by condition $\lim_{m\rightarrow +\infty 
}x_{m}^{}\rightarrow 1$. Then by means of (\ref{8}) we have 
\begin{equation}
x_{m}^{}=1,\qquad m\geq k_{N}^{},  \label{9}
\end{equation}
and
\begin{equation}
x_{m-1}^{}-x_{m}^{}=zq_{m}^{}+o(z),\qquad |z|\rightarrow 0,
\label{31}
\end{equation}
For $m\leq k_{1}^{}-2$ this solution is linear combination of $1$ and 
$z^{m}$. It can be equivalently determined by means of the ``integral" 
equation 
\begin{equation}
x_{m}^{}=1-\sum_{k\geq m}q_{k}^{}\frac{1-z_{}^{k-m}}{1-1/z}x_{k+1}^{}.
\label{14}
\end{equation}
In order to resolve this equation we mention first that due to~(\ref{a2}) it 
can be rewritten as 
\begin{equation}
x_{m}^{}=1-\sum_{j:k_{j}\geq m}\frac{z_{}^{k_{j}-m+1}-z}{1-z}
x_{k_{j}+1}^{}.  \label{17}
\end{equation}
Thus we see that all $x_{m}$ for $m-1\notin {\cal K}$ are determined by
those values for which $m-1\in {\cal K}$. Substituting in (\ref{17}) $m$ by 
$k_{l}+1$, $k_l\in{cal K}$. we get for these values the following linear 
algebraic system 
\begin{equation}
\sum_{j=l}^{N}\frac{z_{}^{k_{j}-k_{l}}-z}{1-z}
x_{k_{j}+1}^{}=1,\qquad l=1,\ldots ,N.  \label{18}
\end{equation}
Let us denote 
\begin{eqnarray}
&&a_{l}^{}(z) =1+\frac{z}{1-z}\sum_{j=l}^{N}x_{k_{j}+1}^{},  \label{19} \\
&&b_{l}^{}(z) =-\frac{1}{1-z}\sum_{j=l}^{N}z_{}^{k_{j}+1}x_{k_{j}+1}^{},
\label{20}
\end{eqnarray}
where $l=1,\ldots ,N$. These functions are not independent as because 
of~(\ref {18}) we have
\begin{equation}
a_{l}^{}+z_{}^{-k_{l}-1}b_{l}^{}=0,\qquad l=1,\ldots ,N,  \label{a22}
\end{equation}
so that (\ref{17}) takes the form
\begin{equation}
x_{m}^{} =\left\{ \begin{array}{ll} 1, & \qquad m\geq k_{N}^{}+1 \\ 
a_{l+1}^{}(1-z^{k_{l+1}-m+1}), & \qquad k_{l+1}^{}\geq m\geq k_{l}^{}+1
\end{array}
\right. , \label{24} 
\end{equation}
where $l=0,\ldots ,N-1$, and we introduced $k_{0}=-\infty$.

On the other side, considering difference $a_{l+1}-a_l$ by~(\ref {19}) we get
\begin{equation}
x_{k_{l}+1}=(1-1/z)(a_{l+1}^{}-a_{l}^{}),\qquad l=1,\ldots ,N-1.  \label{23}
\end{equation}
From~(\ref{19}) at $l=N$ and the upper line of~(\ref {24}) we get that 
\begin{equation}
a_{N}^{}=\frac{1}{1-z}  \label{25}
\end{equation}
and by the second line and (\ref{23}) we derive that
\begin{equation}
a_{l}^{}=a_{l+1}^{}\frac{1-z^{k_{l+1}-k_{l}+1}}{1-z},\qquad
l=1,\ldots ,N-1.  \label{26}
\end{equation}
Thus we get explicitly
\begin{equation}
a_{l}^{}(z)=\frac{1}{1-z}\prod_{j=l}^{N-1}\frac{1-z_{}^{k_{j+1}-k_{j}+1}}
{1-z},\qquad l=1,\ldots ,N,  \label{27}
\end{equation}
and then by (\ref{24}) 
\begin{equation}
x_{m}^{}= \left\{ \begin{array}{ll} 1, & m\geq k_{N}^{} \\ 
\displaystyle\frac{1-z_{}^{k_{l+1}-m+1}}{1-z}\prod_{j=l+1}^{N-1}
\frac{1-z^{k_{j+1}-k_{j}+1}}{1-z}, & k_{l+1}^{}\geq m\geq k_{l}^{}
\end{array} \right.,  \label{28} 
\end{equation}
where $l=0,\ldots ,N-1$, and $k_{0}=-\infty$. In fact we got the first 
line for $m\geq k_{N}+1$. Value $m=k_{N}$ was added as from the second 
line we have that $x_{k_{N}}=1$. In the same way the interval of validity of 
the second equality was extended: we added points $m=k_{l}$ as values at
$m=k_{l}$ and $m=k_{l+1}$ differ by the shift $l\to l+1$ only.

Asymptotic behavior at $w$-infinity, i.e., at $z\to 0$ enables us to 
reconstruct potential, as by~(\ref{14}) we have 
\begin{equation}
x_{m}^{} =1+z\sum_{j=m+1}^{\infty }q_{j}^{}+o(z),\qquad
z\to 0.  \label{32}
\end{equation}
On the other side by (\ref{28}) 
\begin{equation}
x_{m}^{}=\left\{ \begin{array}{ll} 1, & \qquad m\geq k_{N}^{} \\ 
1+z(N-l-\delta _{m,k_{l+1}}^{})+o(z), & \qquad
k_{l+1}^{}\geq m\geq k_{l}^{}
\end{array}
\right.,  \label{a33}
\end{equation}
that coincides with~(\ref {32}).

It was already mentioned above, that only one nontrivial solution exists
in this case, i.e., solution given in~(\ref {28}). Indeed, if we construct
solution $\tilde{x}(z)$ by means of (\ref{13}) we get thanks to~(\ref {10}) 
and~(\ref {28}) relation
\begin{equation}
x_{m}^{}-z_{}^{k_{N}}\tilde{x}_{m}^{}=\theta (m\geq
k_{N}^{})(1-1/z)x_{0,m}^{}.  \label{35}
\end{equation}
It is also impossible to introduce the Jost solution normalized at 
$-\infty$, cf.~(\ref {9}), as solution (\ref{10}) obeys homogeneous equation 
\begin{equation}
x_{0,l}^{}=\sum_{k\leq l}q_{k}^{}\frac{1-z_{}^{k-l}}{1-1/z}
x_{0,k+1}^{}.  \label{16}
\end{equation}

Jost solution~(\ref {28}) of the equation~(\ref {4}) is polynomial with 
respect to $z$ (i.e., $w^{-2}$ because of~(\ref {503})) with coefficients 
belonging to $\N$. It is convenient to introduce new parametrization of this 
solution that will be useful in the construction of the integrals of 
motion below.  So we introduce discrete measure
\begin{equation}
f_{i}^{}(m) =\sum_{n\geq m+1}q_{n}^{}\prod_{j=1}^{i-1}
(1-q_{n+j}^{})q_{n+i}^{}+ \prod_{j=1}^{i-1}(1-q_{m+j}^{})q_{m+i}^{},  
\label{5014},
\end{equation}
where $m\in\Z$ and $i=1,2,\ldots$. Writing down values for the lowest $i$'s
\begin{eqnarray}
f_{1}^{}(m) &=&\sum_{n\geq m+1}q_{n}^{}q_{n+1}^{}+q_{m+1}^{},  \label{5011}\\
f_{2}^{}(m) &=&\sum_{n\geq m+1}q_{n}^{}(1-q_{n+1}^{})q_{n+2}^{}+
(1-q_{m+1}^{})q_{m+2}^{},  \label{5012} \\
f_{3}^{}(m) &=&\sum_{n\geq m+1}q_{n}^{}(1-q_{n+1}^{})(1-q_{n+2}^{})
q_{n+3}^{} +(1-q_{m+1}^{})(1-q_{m+2}^{})q_{m+3}^{},\qquad   \label{5013} 
\end{eqnarray}
etc., we see that for every $m$ and $i$ first term in $f_i(m)$ is equal to 
the number of higher ($i-1$)-defects, i.e., for all $n\geq m+1$ number of 
combinations $\{q_{n},\ldots,q_{n+i}\}=\{1,0,\ldots,0,1\}$ with exactly 
$i-1$ zeroes inside, while the second term is equal to $1$ iff $k_l-m=i$ for 
the lowest $k_l\in{\cal K}$, $k_l\geq m$. In particular, we see that
if $k_{l+1}\geq m\geq k_{l}$ then by~(\ref {5014}) 
\begin{equation}
f_{i}^{}(m)=f_{i}^{}(k_{l+1}^{})+\delta _{k_{l+1}-m,i}.  \label{5021}
\end{equation}
Taking into account that by~(\ref {a2})  $q_{k_{l}}=1$ we get
\begin{equation}
f_{i}^{}(k_{l}^{})=\sum_{n\geq 
k_{l}}q_{n}^{}\prod_{j=1}^{i-1}(1-q_{n+j}^{})q_{n+i}^{},  \label{5017}
\end{equation}
in particular
\begin{equation}
f_{i}^{}(k_{1}^{}) =\sum_{n}q_{n}^{}\prod^{i-1}_{j=1}(1-q_{n+j}^{})
q_{n+i}^{}=\sum_{j\geq l}\delta_{k_{j+1}-k_j,i}^{},\label{5018}
\end{equation}
i.e., $f_{1}(k_1)$ is the number of neighbor cites in the support ${\cal K}$ 
of $q_n$ occupied by units, $f_{i}(k_1)$ are numbers of ($i-1$)-defects in 
the support. As support ${\cal K}$ is bounded (see~(\ref {a2})) we have that
\begin{equation}
\label{5019a}
f_{i}^{}(k_{1}^{}) =0,\qquad i\geq k_N^{}-k_1^{}+2,\qquad l=1,\ldots N,
\end{equation} 
so that by~(\ref {5021})
\begin{equation}
\label{5019b}
f_{i}^{}(m) =\delta _{k_{1}-m,i},\qquad i\geq k_N^{}-k_1^{}+2.
\end{equation}
Also by~(\ref {5014}) it is easy to see that
\begin{eqnarray}
&&f_{i}^{}(m) \equiv 0,\qquad {\rm for}\qquad m\geq k_{N}^{},  \label{5015} \\
&&f_{1}^{}(k_{N}^{}-1) =1,\qquad f_{i}^{}(k_{N}^{}-1)=0,\qquad i\geq 2
\label{5016}
\end{eqnarray}

Definition~(\ref {5014}) can be rewritten in the form
\begin{equation} \label{5020}
f_{i}^{}(m) =\sum_{n\geq m}\prod_{j=1}^{i-1}(1-q_{n+j}^{})q_{n+i}^{}-
\sum_{n\geq m}\prod_{j=1}^{i}(1-q_{n+j}^{})q_{n+i+1}^{} 
\end{equation}
that gives the following relations:
\begin{eqnarray}
\label{50206}
&&\sum_{i=1}^{\infty}f_{i}^{}(m)=\sum_{n\geq m+1}q_n^{}\\
&&\sum_{i=1}^{\infty}if_{i}^{}(m)=\left\{\begin{array}{ll}
0,&\qquad k_N^{}\geq m,\\
k_N^{}-m,&\qquad k_N^{}\leq m,
\end{array}\right. .\label{5026'}
\end{eqnarray}
To prove these equalities, notice first that sums are convergent due 
to~(\ref {5019b}). Then~(\ref {50206}) follows directly from~(\ref {5020}), 
as well as~(\ref {5026'}) for $m\geq k_N^{}$. Let now $k_N^{}-1\geq m$, then
thanks to~(\ref {5020})
$$
\sum_{i=1}^{\infty}if_{i}^{}(m)=\sum_{i=1}^{\infty}
\prod_{j=1}^{i-1}(1-q_{m+j}^{})q_{m+i}^{}=  
\sum_{n\geq m}\Biggl\{\sum_{i=0}^{k_N^{}-1-m}\prod_{j=1}^{i}
(1-q_{n+j}^{})-\sum_{i=0}^{k_N^{}-m}\prod_{j=1}^{i}(1-q_{n+j}^{})\Biggr\}
$$
that proves~(\ref {5026'}).

Let us turn now to~(\ref{28}) for $k_{l+1}^{}\geq m\geq k_{l}^{}$, 
$l=0,\ldots ,N-1$, $k_{0}=-\infty $. Because of~(\ref {5018}) this 
equality can be written for $k_{l+1}^{}>m\geq k_{l}^{}$ as
\[
x_{m}^{}=\prod_{k=1}^{\infty }\left( \frac{1-z_{}^{k+1}}{1-z}\right)
^{\delta _{k_{l+1}-m,k}}\prod_{k=1}^{\infty }\left( \frac{1-z_{}^{k+1}}{1-z}
\right) ^{f_{k}(k_{l+1})} .
\]
Taking~(\ref{5021}) now into account we get finally 
\begin{equation}
x_{m}^{}=\prod_{k=1}^{\infty }\left( \frac{1-z_{}^{k+1}}{1-z}\right)
^{f_{k}(m)}  \label{5033}
\end{equation}
that thanks to~(\ref{5015}) gives also valid result in the first line of 
(\ref{28}) at $m\geq k_{N}$.

\section{Solutions by $\mod 2$}

It was already mentioned that solution~(\ref{5033}) 
is polynomial in $z$ with coefficients belonging to $\N$, so it 
solves~(\ref{8}) by modulo 2, i.e., it obeys
\begin{equation}
x_{m-1}^{}+(1+z)x_{m}^{}+z(1+q_{m}^{})x_{m+1}^{}=0,\qquad \mod 2.  
\label{5031}
\end{equation}
In~[13] it was proved that islands, i.e., intervals
separated by at least three zeroes, evolve independently. Thus it is natural 
to consider here the case where support ${\cal K}$ is just one island. This 
means that we impose condition that all $k_{j}\in {\cal K}$ obey
\begin{equation}
k_{j+1}^{}-k_{j}^{}\leq 3,\qquad j=1,\ldots ,N-1,  \label{37}
\end{equation}
or equivalently,
\begin{equation}
(1-q_{n}^{})(1-q_{n+1}^{})(1-q_{n+2}^{})=0,\qquad 
k_{N}^{}\geq n\geq k_{1}^{}-2.  
\label{5034}
\end{equation}
Then by~(\ref{5014})
\begin{equation}
f_{i}^{}(m)=0,\qquad i\geq 4,\qquad m\geq k_{1}^{}-2,  \label{5035}
\end{equation}
so in the case of only one island our solution~(\ref{5033}) 
is given as product of powers of three polynomials:
\begin{equation}
x_{m}^{}=(1+z)_{}^{f_{1}(m)}(1+z+z_{}^{2})_{}^{f_{2}(m)}(1+z+
z_{}^{2}+z_{}^{3})_{}^{f_{3}(m)},\qquad m\geq k_{1}^{}-2.
\label{502}
\end{equation}
Polynomials $1+z$ and $1+z+z^{2}$ are independent on $\F_2$, but for the 
third polynomial we can write
\begin{equation}
1+z+z_{}^{2}+z_{}^{3}=(1+z)_{}^{3},\qquad \mod 2,  \label{38}
\end{equation}
and because of this
\begin{equation}
x_{m}^{}=(1+z)_{}^{f_{1}(m)+3f_{3}(m)}(1+z+z_{}^{2})_{}^{f_{2}(m)},\qquad 
\mod 2.  \label{504}
\end{equation}
Moreover, by~(\ref{5026'}) and~(\ref{5035}) 
\begin{equation}
f_{1}^{}(m)+2f_{2}^{}(m)+3f_{3}^{}(m)=k_{N}^{}-m,\qquad k_{N}\geq m\geq
k_{1}-2,  \label{506}
\end{equation}
so finally we get
\begin{equation}
x_{m}^{}=(1+z)_{}^{k_{N}-m}(1+z)_{}^{-2f_{2}(m)}(1+z+z_{}^{2})_{}^{f_{2}(m)},
\qquad \mod 2,  \label{510}
\end{equation}
where again $k_{N}\geq m\geq k_{1}-2$. Thus the Jost solution is parametrized
by the only one dynamical variable $f_{2}$. 

The original potential can be reconstructed by means of $f_{2}$. Indeed,
by~(\ref{5012}) 
\[
f_{2}^{}(m-1)-f_{2}^{}(m)=(1-q_{m}^{})(q_{m+1}^{}-(1-q_{m+1}^{})q_{m+2}^{}) 
\]
that thanks to~(\ref{5034}) can be written in the form 
\begin{equation}
f_{2}^{}(m-1)-f_{2}^{}(m)=(1-q_{m}^{})(2q_{m+1}^{}-1),\qquad k_{N}^{}\geq
m\geq k_{1}^{}-2,  \label{5111}
\end{equation} 
and then
\begin{equation}
q_{m}^{}=1-|f_{2}^{}(m)-f_{2}^{}(m-1)|  \label{5112}
\end{equation}
as $|2q_{m+1}-1|=1$. Let us emphasize that equality~(\ref{5112}) is exact, 
i.e., not the $\mod 2$ one.

\section{Time evolution and integrals of motion}

In order to study time evolution by means of the above construction we need 
to consider two neighbor values of time only, say, $t$ and $t+1$. So let us 
simplify notations used in Introduction by denoting
\begin{eqnarray}
&&q_n^{}=q_n^{t},\qquad x_n^{}=x_n^{t},\qquad {\rm etc.}, \label{nohat} \\
&&\hat{q}_n^{}=q_n^{t+1},\qquad \hat{x}_n^{}=x_n^{t+1},\qquad {\rm etc.}
\label{hat}
\end{eqnarray}
Then equation of motion~(\ref {a50}) takes the form
\begin{equation}
\hat{q}_{m}^{}=q_{m}^{}+\hat{q}_{m-2}^{}q_{m+1}^{}+\hat{q}_{m-1}^{}
q_{m+2}^{},\qquad \mod 2,  \label{50}
\end{equation}
and corresponding transformation of the Jost solution is given due 
to~(\ref {a50}) by means of 
\begin{equation}
\hat{\psi}_{m}^{}=\psi_{m}^{}+\hat{q}_{m-1}^{}q_{m+2}^{}\psi_{m+2}^{},\qquad 
\mod 2,
\label{56a}
\end{equation}
or using notations~(\ref {7}) and~(\ref {503}) we have
\begin{equation}
\hat{x}_{m}^{}=x_{m}^{}+z\hat{q}_{m-1}^{}q_{m+2}^{}x_{m+2}^{},\qquad \mod 2.  
\label{56}
\end{equation}
 
Time evolution of $f_2(m)$ is given by substitution of $x_m$ 
in~(\ref{56}) by~(\ref{510}): 
\begin{eqnarray}
&&(1+z)_{}^{2(1-\hat{f}_{2}(m))}(1+z+z_{}^{2})_{}^{\hat{f}_{2}(m)}=
(1+z)_{}^{2(1-f_{2}(m))}(1+z+z_{}^{2})_{}^{f_{2}(m)}+  \nonumber \\
&&\qquad +z\hat{q}_{m-1}^{}q_{m+2}^{}(1+z)_{}^{-2f_{2}(m+2)}
(1+z+z_{}^{2})_{}^{f_{2}(m+2)}, \qquad \mod 2.  \label{67}
\end{eqnarray}
Considering this ($\mod 2$)-equation separately for the cases 
$\hat{q}_{m-1}q_{m+2}$ equal to 0 and to 1, we arrive to the {\bf exact} 
(not $\mod 2$) equation
\begin{equation}
\hat{f}_{2}^{}(m)=f_{2}^{}(m)+\hat{q}_{m-1}^{}q_{m+2}^{}(2q_{m+1}^{}-1),
\qquad m\geq k_{1}-2\label{71}
\end{equation}
($k_{N}\geq m\geq k_{1}-2$ is interval of validity of~(\ref{510}) and 
validity on the interval $m\geq k_{N}+1$ follows from~(\ref{5015})). 
Using~(\ref{5111}) for $f_2$ we get
\begin{eqnarray}
&&\hat{f}_{2}^{}(m-1)-\hat{f}_{2}^{}(m)=  \nonumber \\
&&\qquad =\Bigl(1-q_{m}^{}-\hat{q}_{m-1}^{}q_{m+2}^{}+ \hat{q}
_{m-2}^{}q_{m+1}^{}(2q_{m}^{}-1) \Bigr)(2q_{m+1}^{}-1), \qquad  \label{711}
\end{eqnarray}
and then it is easy to check that $|\hat{f}_{2}(m-1)-\hat{f}_{2}(m)|$ is
equal to $0$ or $1$, so we can use~(\ref{5112}) to define 
\begin{equation}
\hat{q}_{m}^{}=1-|\hat{f}_{2}^{}(m-1)-\hat{f}_{2}^{}(m)|.  \label{73}
\end{equation}
Thanks to this relation we derive the exact equation of time evolution of
$q_m$, that substitutes ($\mod 2$)-equation~(\ref {50}). Indeed, 
using~(\ref{711}) we get 
\begin{equation}
\hat{q}_{m}^{}=1-|1-q_{m}^{}-\hat{q}_{m-1}^{}q_{m+2}^{}+\hat{q}
_{m-2}^{}q_{m+1}^{}(2q_{m}^{}-1)|.  \label{76}
\end{equation}
The r.h.s.\ can be simplified if we notice that 
$$
|1-q_{m}^{}-\hat{q}_{m-1}^{}q_{m+2}^{}+\hat{q}
_{m-2}^{}q_{m+1}^{}(2q_{m}^{}-1)|=
1-q_{m}^{}+|\hat{q}_{m-1}^{}q_{m+2}^{}-\hat{q}_{m-2}^{}q_{m+1}^{}|
(2q_{m}^{}-1),
$$
so finally we get equation
\begin{equation}
\hat{q}_{m}^{}=q_{m}^{}-|\hat{q}_{m-1}^{}q_{m+2}^{}-\hat{q}
_{m-2}^{}q_{m+1}^{}|(2q_{m}^{}-1).  \label{79}
\end{equation}

In order to consider integrals of motion we introduce elements of the 
monodromi matrix by means of the standard relation
\begin{equation}
x_{m}^{}=a(z)+z_{}^{-m}b(z),\qquad m\le k_{1}^{},  \label{101}
\end{equation}
and analogously for $\hat{x}_{m}^{}$. Thanks to (\ref{56}) for $m\le k_{1}$
$$
\hat{x}_{m}^{}=x_{m}^{},  \label{102}
$$
thus
\begin{equation}
\hat{a}(z)=a(z),\qquad \hat{b}(z)=b(z)  \label{103}
\end{equation}
i.e., both $a(z)$ and $b(z)$ are integrals of motion as it was demonstrated 
in~[13]. Here because of~(\ref{510}) we single out functionally independent 
integrals explicitly. Indeed, by~(\ref{24}) 
\begin{equation}
b(z)=z^{1+k_{1}}_{}a(z),\qquad (1-z)a(z)=x_{k_{1}}^{}(z),\qquad \mod 2,
\label{104}
\end{equation}
so $x_{k_1}(z)$ is also integral of motion and by~(\ref{510}) we have
\begin{equation}
x_{k_1}^{}(z)=(1+z)_{}^{k_{N}-k_{1}}(1+z)_{}^{-2f_{2}(k_{1})}
(1+z+z_{}^{2})_{}^{f_{2}(k_{1})},\qquad 
\mod 2.  \label{105}
\end{equation}
Now~(\ref{103}) and~(\ref{104}) are equivalent to
\begin{eqnarray}
&&\hat{k}_1^{}=k_1^{},\qquad \hat{k}_{\hat N}^{}=k_N^{},\nonumber\\
&&\hat{f}_{2}(k_{1})=f_{2}(k_{1}),  \label{106}
\end{eqnarray}
where the first line was already known (see~(\ref{a63})). 

Thus in addition to two known integrals---borders of the island, $k_1$ and 
$k_N$,---we proved that $f_2(k_1)$, number of the single defects (isolated 
zeroes, see comments after~(\ref{5018})) inside the island, is preserved 
under time evolution. This new integral is explicitly demonstrated on
Fig.\ 1. It is clear that these three integrals are mutually independent and 
thanks to~(\ref{104}) and~(\ref{105}) they uniquely determine $a(z)$ and 
$b(z)$. Thus all coefficients of the Taylor expansion of $a(z)$ are 
functions of these three integrals only. We emphasize that in
spite of the fact that Eq.~(\ref{105}) is equation by $\mod 2$ all
these three integrals are exact ones. 

An example of ($\mod 2$)-integral is provided by $N$. Indeed, in analogy 
with~(\ref{506}) we have from~(\ref{50206}) that
$$
f_{1}^{}(m)+f_{2}^{}(m)+f_{3}^{}(m)=\sum_{j\geq m+1}q_{j}^{}.
$$
On the other side~(\ref{506}) gives
$$
f_{1}^{}(m)+f_{3}^{}(m)=k_{N}^{}-m,\qquad \mod 2,
$$
so that
$$
\sum_{j\geq m+1}q_{j}^{}=k_{N}^{}-m+f_{2}^{}(m),\qquad \mod 2.
$$
Then
$$
N=\sum_{m\geq k_1+1}q_{m}^{}+1=k_{N}^{}-k_{1}^{}+1+f_{2}(k_{1}),\qquad \mod 2,
$$
and as any exact integral of motion is also ($\mod 2$)-integral we obtain
that
\begin{equation}
\hat{N}=N,\qquad \mod 2.  \label{121}
\end{equation}

Explicit formulas for $f_1(k_1)$, $f_2(k_1)$, and $f_3(k_1)$ are given 
in~(\ref{5018}) and Fig.~1  shows that indeed neither $f_1(k_1)$, nor 
$f_3(k_1)$ are preserved under time evolution. Of course, this does not mean
that there cannot be some other integrals of motion. Moreover, it is unclear 
if existence of three exact integrals~(\ref{106}) and one 
($\mod 2$)-integral~(\ref{121}) is enough for the system to be integrable in 
some sense. 

One of results of the above construction is substitution of ($\mod 
2$)-equation~(\ref{a50}) (or~(\ref{50})) by~(\ref{79}). The latter is the 
exact one, as for $q_m$ with range equal $\{0,1\}$ we get $\hat{q}_m$ with 
the same range $\{0,1\}$ without 
using summation by $\mod 2$. Thus the problem of integrability of this
cellular automaton is reduced to the problem of existence
of an (exact!) Lax pair for the equation~(\ref {79}). This problem, as well 
as problem of definition of the reasonable spectral data for the linear 
operator~(\ref{a50}) on $\F_2$ are still opened. Moreover, we have to 
stress, that such notions as ``integrability" and ``degree of freedom" 
themselves must be modified to be applicable to the case of systems on finite 
fields.

{\bf Acknowledgments.} Work is supported in part by RFBR under Grant
No.~96-01-00344 and by INTAS under Grant No.~93-166-ext. Author thanks 
M.~Bruschi and O.~Ragnisco for fruitful discussions.

\pagebreak
\baselineskip2pt

$\scriptstyle\sq\sq\sq\bsq \bsq \bsq \sq \bsq \bsq \bsq \bsq \bsq \sq 
\sq \bsq \bsq \bsq \bsq \sq \sq \sq \bsq \bsq \sq \sq \bsq \bsq \bsq \bsq \sq 
\bsq \bsq \bsq \bsq \bsq \bsq \sq \sq \bsq \bsq \sq \bsq \bsq \sq \sq \bsq 
\bsq \bsq \bsq\sq\sq\sq $\nopagebreak

$\scriptstyle\sq\sq\sq\bsq \bsq \sq \bsq \sq \sq \bsq \bsq \bsq \bsq \sq 
\sq \bsq \bsq \bsq \sq \sq \sq \bsq \bsq \bsq \sq \sq \bsq \bsq \sq \bsq \sq 
\sq \bsq \sq \sq \bsq \bsq \sq \sq \bsq \bsq \sq \bsq \bsq \sq \sq \bsq \bsq 
\bsq\sq\sq\sq $\nopagebreak

$\scriptstyle\sq\sq\sq\bsq \sq \bsq \bsq \bsq \sq \sq \bsq \bsq \bsq 
\bsq \sq \sq \bsq \bsq \sq \sq \sq \bsq \bsq \bsq \bsq \sq \sq \bsq \sq \bsq 
\bsq \bsq \bsq \bsq \sq \sq \bsq \bsq \sq \sq \bsq \bsq \sq \bsq \bsq \sq \sq 
\bsq \bsq\sq\sq\sq $\nopagebreak

$\scriptstyle\sq\sq\sq\bsq \bsq \bsq \sq \bsq \bsq \sq \sq \bsq \bsq 
\bsq \bsq \sq \sq \bsq \sq \sq \sq \bsq \sq \sq \bsq \bsq \bsq \sq \bsq \sq 
\sq \bsq \bsq \bsq \bsq \sq \sq \bsq \bsq \sq \sq \bsq \bsq \sq \bsq \bsq \sq 
\sq \bsq\sq\sq\sq $\nopagebreak 

$\scriptstyle\sq\sq\sq\bsq \bsq \sq \bsq \bsq \bsq \bsq \sq \sq \bsq 
\bsq \bsq \bsq \bsq \bsq \sq \sq \sq \bsq \bsq \sq \sq \bsq \sq \bsq \bsq 
\bsq 
\sq \sq \bsq \bsq \bsq \bsq \sq \sq \bsq \bsq \sq \sq \bsq \bsq \sq \bsq \bsq 
\bsq \bsq\sq\sq\sq $\nopagebreak

$\scriptstyle\sq\sq\sq\bsq \sq \bsq \sq \sq \bsq \bsq \bsq \sq \sq \bsq 
\sq \sq \bsq \bsq \sq \sq \sq \bsq \bsq \bsq \bsq \sq \bsq \sq \sq \bsq \bsq 
\sq \sq \bsq \bsq \bsq \bsq \sq \sq \bsq \bsq \sq \sq \bsq \bsq \bsq \bsq \sq 
\bsq\sq\sq\sq $\nopagebreak

$\scriptstyle\sq\sq\sq\bsq \sq \bsq \bsq \sq \sq \bsq \bsq \bsq \bsq 
\bsq \bsq \sq \sq \bsq \sq \sq \sq \bsq \sq \sq \bsq \sq \bsq \bsq \sq \sq 
\bsq \bsq \sq \sq \bsq \bsq \bsq \bsq \sq \sq \bsq \bsq \sq \sq \bsq \bsq \sq 
\bsq \bsq\sq\sq\sq $\nopagebreak

$\scriptstyle\sq\sq\sq\bsq \bsq \sq \bsq \bsq \sq \sq \bsq \sq \sq \bsq 
\bsq \bsq \bsq \bsq \sq \sq \sq \bsq \bsq \bsq \sq \bsq \bsq \bsq \bsq \sq 
\sq 
\bsq \bsq \sq \sq \bsq \bsq \bsq \bsq \sq \sq \bsq \bsq \sq \sq \bsq \bsq \sq 
\bsq\sq\sq\sq $\nopagebreak

$\scriptstyle\sq\sq\sq\bsq \sq \bsq \bsq \bsq \bsq \bsq \bsq \bsq \sq 
\sq \bsq \sq \sq \bsq \sq \sq \sq \bsq \bsq \sq \bsq \sq \sq \bsq \bsq \bsq 
\sq \sq \bsq \bsq \sq \sq \bsq \bsq \bsq \bsq \sq \sq \bsq \bsq \sq \sq \bsq 
\sq \bsq\sq\sq\sq $\nopagebreak

$\scriptstyle\sq\sq\sq\bsq \bsq \bsq \bsq \bsq \bsq \bsq \sq \bsq \bsq 
\bsq \bsq \bsq \bsq \bsq \sq \sq \sq \bsq \sq \bsq \bsq \bsq \sq \sq \bsq 
\bsq 
\bsq \sq \sq \bsq \bsq \sq \sq \bsq \bsq \bsq \bsq \sq \sq \bsq \bsq \bsq \sq 
\bsq \bsq\sq\sq\sq $\nopagebreak

$\scriptstyle\sq\sq\sq\bsq \sq \sq \bsq \sq \sq \bsq \bsq \bsq \bsq \bsq 
\bsq \bsq \sq \bsq \sq \sq \sq \bsq \bsq \bsq \sq \bsq \bsq \sq \sq \bsq \bsq 
\bsq \sq \sq \bsq \bsq \sq \sq \bsq \bsq \bsq \bsq \sq \sq \bsq \sq \bsq \bsq 
\bsq\sq\sq\sq $\nopagebreak

$\scriptstyle\sq\sq\sq\bsq \bsq \bsq \bsq \bsq \sq \sq \bsq \sq \sq \bsq 
\bsq \sq \bsq \bsq \sq \sq \sq \bsq \bsq \sq \bsq \bsq \bsq \bsq \sq \sq \bsq 
\bsq \bsq \sq \sq \bsq \bsq \sq \sq \bsq \bsq \bsq \bsq \bsq \sq \bsq \sq \sq 
\bsq\sq\sq\sq $\nopagebreak

$\scriptstyle\sq\sq\sq\bsq \sq \sq \bsq \bsq \bsq \bsq \bsq \bsq \sq \sq 
\bsq \bsq \sq \bsq \sq \sq \sq \bsq \sq \bsq \sq \sq \bsq \bsq \bsq \sq \sq 
\bsq \bsq \bsq \sq \sq \bsq \bsq \sq \sq \bsq \sq \sq \bsq \sq \bsq \bsq \bsq 
\bsq\sq\sq\sq $\nopagebreak

$\scriptstyle\sq\sq\sq\bsq \bsq \sq \sq \bsq \sq \sq \bsq \bsq \bsq \sq 
\sq \bsq \sq \bsq \sq \sq \sq \bsq \sq \bsq \bsq \sq \sq \bsq \bsq \bsq \sq 
\sq \bsq \bsq \bsq \sq \sq \bsq \bsq \bsq \bsq \bsq \bsq \sq \bsq \sq \sq 
\bsq 
\bsq\sq\sq\sq $\nopagebreak

$\scriptstyle\sq\sq\sq\bsq \bsq \bsq \bsq \bsq \bsq \sq \sq \bsq \bsq 
\bsq \bsq \sq \bsq \bsq \sq \sq \sq \bsq \bsq \sq \bsq \bsq \sq \sq \bsq \bsq 
\bsq \sq \sq \bsq \bsq \bsq \sq \sq \bsq \sq \sq \bsq \sq \bsq \bsq \bsq \sq 
\sq \bsq\sq\sq\sq $\nopagebreak

$\scriptstyle\sq\sq\sq\bsq \sq \sq \bsq \bsq \bsq \bsq \sq \sq \bsq \bsq 
\sq \bsq \bsq \bsq \sq \sq \sq \bsq \sq \bsq \bsq \bsq \bsq \sq \sq \bsq \bsq 
\bsq \sq \sq \bsq \bsq \bsq \bsq \bsq \bsq \bsq \sq \bsq \sq \sq \bsq \bsq 
\bsq 
\bsq\sq\sq\sq $\nopagebreak

$\scriptstyle\sq\sq\sq\bsq \bsq \sq \sq \bsq \bsq \bsq \bsq \sq \sq \bsq 
\bsq \bsq \sq \bsq \sq \sq \sq \bsq \bsq \bsq \bsq \sq \bsq \bsq \sq \sq \bsq 
\bsq \bsq \sq \sq \bsq \sq \sq \bsq \bsq \sq \bsq \bsq \bsq \sq \sq \bsq \bsq 
\bsq\sq\sq\sq $\nopagebreak

$\scriptstyle\sq\sq\sq\bsq \bsq \bsq \sq \sq \bsq \bsq \bsq \bsq \sq \sq 
\bsq \sq \bsq \bsq \sq \sq \sq \bsq \sq \sq \bsq \bsq \sq \bsq \bsq \sq \sq 
\bsq \bsq \bsq \bsq \bsq \bsq \sq \sq \bsq \bsq \bsq \sq \bsq \bsq \sq \sq 
\bsq 
\bsq\sq\sq\sq $\nopagebreak

$\scriptstyle\sq\sq\sq\bsq \bsq \bsq \bsq \sq \sq \bsq \bsq \bsq \bsq 
\bsq \sq \bsq \bsq \bsq \sq \sq \sq \bsq \bsq \sq \sq \bsq \bsq \sq \bsq \bsq 
\sq \sq \bsq \sq \sq \bsq \bsq \bsq \sq \sq \bsq \sq \bsq \bsq \bsq \bsq \sq 
\sq \bsq\sq\sq\sq $\nopagebreak

$\scriptstyle\sq\sq\sq\bsq \sq \sq \bsq \bsq \sq \sq \bsq \sq \sq \bsq 
\bsq \bsq \sq \bsq \sq \sq \sq \bsq \bsq \bsq \sq \sq \bsq \bsq \sq \bsq \bsq 
\bsq \bsq \bsq \sq \sq \bsq \bsq \bsq \bsq \sq \bsq \sq \sq \bsq \bsq \bsq 
\bsq 
\bsq\sq\sq\sq $\nopagebreak

$\scriptstyle\sq\sq\sq\bsq \bsq \sq \sq \bsq \bsq \bsq \bsq \bsq \sq \sq 
\bsq \sq \bsq \bsq \sq \sq \sq \bsq \bsq \bsq \bsq \sq \sq \bsq \bsq \bsq 
\bsq 
\bsq \sq \bsq \bsq \sq \sq \bsq \bsq \sq \bsq \bsq \bsq \sq \sq \bsq \sq \sq 
\bsq\sq\sq\sq $\nopagebreak

$\scriptstyle\sq\sq\sq\bsq \bsq \bsq \sq \sq \bsq \sq \sq \bsq \bsq \bsq 
\sq \bsq \bsq \bsq \sq \sq \sq \bsq \sq \sq \bsq \bsq \sq \sq \bsq \sq \sq 
\bsq \bsq \sq \bsq \bsq \sq \sq \bsq \bsq \bsq \sq \bsq \bsq \bsq \bsq \bsq 
\bsq \bsq\sq\sq\sq $\nopagebreak

$\scriptstyle\sq\sq\sq\bsq \bsq \bsq \bsq \bsq \bsq \bsq \sq \sq \bsq 
\sq \bsq \sq \sq \bsq \sq \sq \sq \bsq \bsq \sq \sq \bsq \bsq \bsq \bsq \bsq 
\sq \sq \bsq \bsq \sq \bsq \bsq \sq \sq \bsq \sq \bsq \sq \sq \bsq \sq \sq 
\bsq 
\bsq\sq\sq\sq $\nopagebreak

$\scriptstyle\sq\sq\sq\bsq \sq \sq \bsq \sq \sq \bsq \bsq \bsq \sq \bsq 
\bsq \bsq \bsq \bsq \sq \sq \sq \bsq \bsq \bsq \sq \sq \bsq \sq \sq \bsq \bsq 
\sq \sq \bsq \bsq \sq \bsq \bsq \bsq \sq \bsq \bsq \bsq \bsq \bsq \bsq \sq 
\sq 
\bsq\sq\sq\sq $\nopagebreak 

$\scriptstyle\sq\sq\sq\bsq \bsq \bsq \bsq \bsq \sq \sq \bsq \sq \bsq \sq 
\sq \bsq \bsq \bsq \sq \sq \sq \bsq \bsq \bsq \bsq \bsq \bsq \bsq \sq \sq 
\bsq 
\bsq \sq \sq \bsq \bsq \bsq \sq \bsq \bsq \bsq \bsq \bsq \bsq \sq \bsq \bsq 
\bsq \bsq\sq\sq\sq $\nopagebreak

$\scriptstyle\sq\sq\sq\bsq \sq \sq \bsq \bsq \bsq \bsq \sq \bsq \bsq 
\bsq \sq \sq \bsq \bsq \sq \sq \sq \bsq \sq \sq \bsq \sq \sq \bsq \bsq \sq 
\sq 
\bsq \bsq \sq \sq \bsq \sq \bsq \sq \sq \bsq \sq \sq \bsq \bsq \bsq \bsq \sq 
\bsq\sq\sq\sq $\nopagebreak

$\scriptstyle\sq\sq\sq\bsq \bsq \sq \sq \bsq \bsq \sq \bsq \sq \sq \bsq 
\bsq \sq \sq \bsq \sq \sq \sq \bsq \bsq \bsq \bsq \bsq \sq \sq \bsq \bsq \sq 
\sq \bsq \bsq \bsq \sq \bsq \bsq \bsq \bsq \bsq \bsq \sq \sq \bsq \bsq \sq 
\bsq 
\bsq\sq\sq\sq $\nopagebreak

$\scriptstyle\sq\sq\sq\bsq \bsq \bsq \sq \sq \bsq \sq \bsq \bsq \sq \sq 
\bsq \bsq \bsq \bsq \sq \sq \sq \bsq \sq \sq \bsq \bsq \bsq \sq \sq \bsq \bsq 
\sq \sq \bsq \sq \bsq \sq \sq \bsq \sq \sq \bsq \bsq \sq \sq \bsq \bsq \sq 
\bsq\sq\sq\sq $\nopagebreak 

$\scriptstyle\sq\sq\sq\bsq \bsq \bsq \bsq \bsq \sq \bsq \bsq \bsq \bsq 
\sq \sq \bsq \bsq \bsq \sq \sq \sq \bsq \bsq \sq \sq \bsq \bsq \bsq \sq \sq 
\bsq \bsq \bsq \sq \bsq \bsq \bsq \bsq \bsq \bsq \sq \sq \bsq \bsq \sq \sq 
\bsq \sq \bsq\sq\sq\sq $\nopagebreak

$\scriptstyle\sq\sq\sq\bsq \sq \sq \bsq \sq \bsq \sq \sq \bsq \bsq \bsq 
\sq \sq \bsq \bsq \sq \sq \sq \bsq \bsq \bsq \sq \sq \bsq \bsq \bsq \sq \sq 
\bsq \sq \bsq \sq \sq \bsq \sq \sq \bsq \bsq \sq \sq \bsq \bsq \bsq \sq \bsq 
\bsq\sq\sq\sq $\nopagebreak

$\scriptstyle\sq\sq\sq\bsq \bsq \bsq \sq \bsq \bsq \bsq \sq \sq \bsq 
\bsq \bsq \sq \sq \bsq \sq \sq \sq \bsq \bsq \bsq \bsq \sq \sq \bsq \bsq \bsq 
\bsq \sq \bsq \bsq \bsq \bsq \bsq \bsq \sq \sq \bsq \bsq \sq \sq \bsq \sq 
\bsq 
\bsq \bsq\sq\sq\sq $\nopagebreak

$\scriptstyle\sq\sq\sq\bsq \bsq \sq \bsq \sq \sq \bsq \bsq \sq \sq \bsq 
\bsq \bsq \bsq \bsq \sq \sq \sq \bsq \sq \sq \bsq \bsq \sq \sq \bsq \bsq \sq 
\bsq \sq \sq \bsq \sq \sq \bsq \bsq \sq \sq \bsq \bsq \bsq \sq \bsq \sq \sq 
\bsq\sq\sq\sq $\nopagebreak

$\scriptstyle\sq\sq\sq\bsq \sq \bsq \bsq \bsq \sq \sq \bsq \bsq \sq \sq 
\bsq \sq \sq \bsq \sq \sq \sq \bsq \bsq \sq \sq \bsq \bsq \sq \sq \bsq \sq 
\bsq \bsq \bsq \bsq \bsq \sq \sq \bsq \bsq \sq \sq \bsq \sq \bsq \bsq \bsq 
\bsq 
\bsq\sq\sq\sq $\nopagebreak

$\scriptstyle\sq\sq\sq\bsq \bsq \bsq \sq \bsq \bsq \sq \sq \bsq \bsq 
\bsq \bsq \bsq \bsq \bsq \sq \sq \sq \bsq \bsq \bsq \sq \sq \bsq \bsq \bsq 
\sq 
\bsq \sq \sq \bsq \bsq \bsq \bsq \sq \sq \bsq \bsq \bsq \sq \bsq \sq \sq \bsq 
\bsq \bsq\sq\sq\sq $\nopagebreak

$\scriptstyle\sq\sq\sq\bsq \bsq \sq \bsq \bsq \bsq \bsq \sq \sq \bsq \sq 
\sq \bsq \bsq \bsq \sq \sq \sq \bsq \bsq \bsq \bsq \sq \sq \bsq \sq \bsq \bsq 
\bsq \sq \sq \bsq \bsq \bsq \bsq \sq \sq \bsq \sq \bsq \bsq \bsq \sq \sq \bsq 
\bsq\sq\sq\sq $\nopagebreak

$\scriptstyle\sq\sq\sq\bsq \sq \bsq \sq \sq \bsq \bsq \bsq \bsq \bsq 
\bsq \sq \sq \bsq \bsq \sq \sq \sq \bsq \sq \sq \bsq \bsq \bsq \sq \bsq \sq 
\sq \bsq \bsq \sq \sq \bsq \bsq \bsq \bsq \bsq \sq \bsq \sq \sq \bsq \bsq \sq 
\sq \bsq\sq\sq\sq $\nopagebreak

$\scriptstyle\sq\sq\sq\bsq \sq \bsq \bsq \sq \sq \bsq \sq \sq \bsq \bsq 
\bsq \sq \sq \bsq \sq \sq \sq \bsq \bsq \sq \sq \bsq \sq \bsq \bsq \bsq \sq 
\sq \bsq \bsq \sq \sq \bsq \sq \sq \bsq \sq \bsq \bsq \sq \sq \bsq \bsq \bsq 
\bsq\sq\sq\sq $\nopagebreak

$\scriptstyle\sq\sq\sq\bsq \bsq \sq \bsq \bsq \bsq \bsq \bsq \sq \sq 
\bsq \bsq \bsq \bsq \bsq \sq \sq \sq \bsq \bsq \bsq \bsq \sq \bsq \sq \sq 
\bsq 
\bsq \sq \sq \bsq \bsq \bsq \bsq \bsq \bsq \sq \bsq \bsq \bsq \bsq \sq \sq 
\bsq \bsq \bsq\sq\sq\sq $\nopagebreak

$\scriptstyle\sq\sq\sq\bsq \sq \bsq \sq \sq \bsq \bsq \bsq \bsq \sq \sq 
\bsq \sq \sq \bsq \sq \sq \sq \bsq \sq \sq \bsq \sq \bsq \bsq \sq \sq \bsq 
\bsq \sq \sq \bsq \sq \sq \bsq \sq \bsq \sq \sq \bsq \bsq \bsq \sq \sq \bsq 
\bsq\sq\sq\sq $\nopagebreak

$\scriptstyle\sq\sq\sq\bsq \sq \bsq \bsq \sq \sq \bsq \bsq \bsq \bsq 
\bsq \bsq \bsq \bsq \bsq \sq \sq \sq \bsq \bsq \bsq \sq \bsq \bsq \bsq \bsq 
\sq \sq \bsq \bsq \bsq \bsq \bsq \bsq \sq \bsq \bsq \bsq \sq \sq \bsq \bsq 
\bsq \sq \sq \bsq\sq\sq\sq $\nopagebreak

$\scriptstyle\sq\sq\sq\bsq \bsq \sq \bsq \bsq \sq \sq \bsq \sq \sq \bsq 
\sq \sq \bsq \bsq \sq \sq \sq \bsq \bsq \sq \bsq \sq \sq \bsq \bsq \bsq \sq 
\sq \bsq \sq \sq \bsq \sq \bsq \sq \sq \bsq \bsq \sq \sq \bsq \bsq \bsq \bsq 
\bsq\sq\sq\sq $\nopagebreak

$\scriptstyle\sq\sq\sq\bsq \sq \bsq \bsq \bsq \bsq \bsq \bsq \bsq \bsq 
\bsq \bsq \sq \sq \bsq \sq \sq \sq \bsq \sq \bsq \bsq \bsq \sq \sq \bsq \bsq 
\bsq \bsq \bsq \bsq \bsq \sq \bsq \bsq \bsq \sq \sq \bsq \bsq \sq \sq \bsq 
\sq 
\sq \bsq\sq\sq\sq $\nopagebreak

$\scriptstyle\sq\sq\sq\bsq \bsq \bsq \bsq \bsq \bsq \bsq \bsq \bsq \bsq 
\sq \bsq \bsq \bsq \bsq \sq \sq \sq \bsq \bsq \bsq \sq \bsq \bsq \sq \sq \bsq 
\sq \sq \bsq \bsq \sq \bsq \sq \sq \bsq \bsq \sq \sq \bsq \bsq \bsq \bsq \bsq 
\bsq \bsq\sq\sq\sq $\nopagebreak

$\scriptstyle\sq\sq\sq\bsq \sq \sq \bsq \sq \sq \bsq \sq \sq \bsq \bsq 
\bsq \bsq \sq \bsq \sq \sq \sq \bsq \bsq \sq \bsq \bsq \bsq \bsq \bsq \bsq 
\bsq \sq \sq \bsq \sq \bsq \bsq \sq \sq \bsq \bsq \sq \sq \bsq \sq \sq \bsq 
\bsq \bsq\sq\sq\sq $\nopagebreak

$\scriptstyle\sq\sq\sq\bsq \bsq \bsq \bsq \bsq \bsq \bsq \bsq \sq \sq 
\bsq \bsq \sq \bsq \bsq \sq \sq \sq \bsq \sq \bsq \sq \sq \bsq \sq \sq \bsq 
\bsq \bsq \bsq \sq \bsq \bsq \bsq \bsq \sq \sq \bsq \bsq \bsq \bsq \bsq \sq 
\sq \bsq \bsq\sq\sq\sq $\nopagebreak

$\scriptstyle\sq\sq\sq\bsq \sq \sq \bsq \sq \sq \bsq \bsq \bsq \sq \sq 
\bsq \bsq \sq \bsq \sq \sq \sq \bsq \sq \bsq \bsq \bsq \bsq \bsq \sq \sq \bsq 
\bsq \sq \bsq \sq \sq \bsq \bsq \bsq \sq \sq \bsq \sq \sq \bsq \bsq \sq \sq 
\bsq\sq\sq\sq $\nopagebreak

$\scriptstyle\sq\sq\sq\bsq \bsq \bsq \bsq \bsq \sq \sq \bsq \bsq \bsq 
\sq \sq \bsq \sq \bsq \sq \sq \sq \bsq \bsq \bsq \bsq \bsq \sq \bsq \bsq \sq 
\sq \bsq \sq \bsq \bsq \sq \sq \bsq \bsq \bsq \bsq \bsq \bsq \sq \sq \bsq 
\bsq 
\bsq \bsq\sq\sq\sq $\nopagebreak
\vspace{1cm}

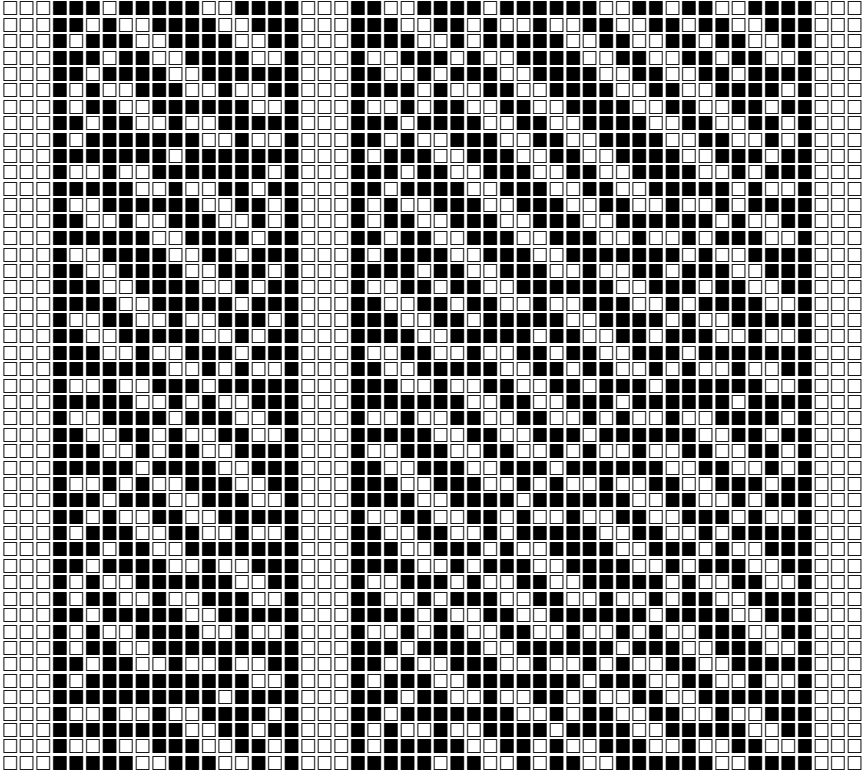
\centerline{{\bf Fig.~1.} Here the horizontal axis is the $n$-axis and the 
downward vertical axis is the $t$-axis.}

\end{document}